\documentclass[journal]{IEEEtran}
\usepackage{graphicx,amssymb,amsmath}
\usepackage{multicol}

\usepackage{setspace}
\usepackage{stfloats}
\usepackage{amsthm}
\usepackage{flushend}
\usepackage{cases,subeqnarray}
\usepackage{bm,multirow,bigstrut}
\usepackage{textcomp}
\usepackage{latexsym,bm}
\usepackage{booktabs,changebar}
\usepackage{xcolor}
\usepackage{mathtools}
\usepackage{dsfont}
\usepackage{extarrows}
\usepackage{cite}

\theoremstyle{plain}

\newtheorem{lemm}{Lemma}
\theoremstyle{plain}

\IEEEoverridecommandlockouts
\begin{document}
\title{Physical Layer Security over Fluctuating Two-Ray Fading Channels}
\author{Wen Zeng, Jiayi~Zhang,~\IEEEmembership{Member,~IEEE,} Shuaifei Chen, Kostas P. Peppas,~\IEEEmembership{Senior Member,~IEEE,} \\
and Bo Ai,~\IEEEmembership{Senior Member,~IEEE}

\thanks{Copyright (c) 2015 IEEE. Personal use of this material is permitted. However, permission to use this material for any other purposes must be obtained from the IEEE by sending a request to pubs-permissions@ieee.org.}
\thanks{W. Zeng, J. Zhang and S. Chen are with the School of Electronic and Information Engineering, Beijing Jiaotong University, Beijing 100044, China. 
(e-mail: jiayizhang@bjtu.edu.cn).}
\thanks{K. P. Peppas is with the Department of Informatics and Telecommunications, University of Peloponnese, 22100 Tripoli, Greece (e-mail: peppas@uop.gr).}
\thanks{B. Ai is with the State Key Laboratory of Rail Traffic Control and Safety, Beijing Jiaotong University, Beijing 100044, China (e-mail: aibo@ieee.org).}
}

\maketitle

\begin{abstract}
Ensuring the physical layer security (PHY-security) of millimeter wave (mmWave) communications is one of the key factors for the success of 5G. Recent field measurements show that conventional fading models cannot accurately model the random fluctuations of mmWave signals. To tackle this challenge, the fluctuating two-ray (FTR) fading model has been proposed. In this correspondence, we comprehensively analyze the PHY-security in mmWave communications over FTR fading channels. More specifically, we derive analytical expressions for significant PHY-security metrics, such as average secrecy capacity, secrecy outage probability, and the probability of strictly positive secrecy capacity, with simple functions. The effect of channel parameters on the PHY-security has been validated by numerical results.
\end{abstract}

\IEEEpeerreviewmaketitle

\section{Introduction}
As a promising technique for supporting skyrocket data rate in fifth-generation (5G), millimeter wave (mmWave) communications have received an increasing attention due to the large available bandwidth at mmWave frequencies \cite{zhang2018low}. Given the ubiquitousness of wireless channels, mmWave communications are particularly vulnerable to a set of eavesdropping and impersonation attacks. Compared to cryptographic technologies implemented at upper layers, physical layer security (PHY-security) is a low-complexity alternative that exploits the randomness of wireless channels to safeguard the confidential information transmission \cite{Liu2017Physical}.

An increasing number of literatures show their interests of exploring the PHY-security in mmWave communications \cite{Zhu2017secure,Ju2017secure,wang2016physical}. For example, the effect of peculiar mmWave channel characteristics on the PHY-security performance in mmWave Ad hoc networks has been studied in \cite{Zhu2017secure}. In \cite{Ju2017secure}, PHY-security transmissions under slow fading channels with multipath propagation in mmWave communications were studied. However, both \cite{Zhu2017secure} and \cite{Ju2017secure} neglected the small-scale fading of mmWave channel. Leveraging on a stochastic geometry framework, the authors of \cite{wang2016physical} investigated the downlink PHY-security performance in an mmWave cellular network assuming Nakagami-$m$ fading. Moreover, the PHY-security performance of hybrid mmWave networks has been investigated in \cite{Vuppala2016An,Vuppala2017On}.

Most of works only pay attention to the PHY-security in mmWave communications over slow fading channels. The small-scale channel model is also important for taking a deeper look into signal processing for mmWave communications, such as beamforming and precoding. Very recently, a 28 GHz outdoor measurement campaign showed that conventional small-scale fading models \cite{Zhang2017Performance} (e.g., Rayleigh, Rician and Nakagami-$m$) cannot accurately model the random fluctuations suffered by mmWave signals \cite{sun2017millimeter}. In order to circumvent this issue, the fluctuating two-ray (FTR) fading model proposed in  \cite{Jerez2017the} can capture the bimodality of mmWave channels, which is more accurate than conventional fading models.

Therefore, the PHY-security performance of mmWave communications over FTR fading channels is still a significant and unsolved problem. Motivated by that, we provide a further investigation on the comprehensive analysis of the PHY-security performance of mmWave communications and derive analytical exact expressions for the average secrecy capacity (ASC), the secrecy outage probability (SOP), and the probability of strictly positive secrecy capacity (SPSC).
Since the FTR includes Rayleigh, Rician, and Nakagami-$m$ as special cases, the derived results can reduce to many pioneering works. Moreover, our work is beneficial to evaluate the state of the art PHY-security techniques and get better insight into the application of the FTR fading models in practical mmWave communications.

\section{System Model}\label{se:model}
Hereafter, we consider the classic Wyner's wiretap model, which has been widely applied in the PHY-security analysis for mmWave communications \cite{wang2016physical,Vuppala2016An,Vuppala2017On,Wang2014Physical}. Suppose that the source $S$ sends a message to the legitimate receiver $D$ over the main channel while the eavesdropper $E$ attempts to decode this message from its received signal through the eavesdropper channel. It is assumed that the main and eavesdropper channels experience independent FTR fading. Furthermore, we assume that the full channel state information (CSI) of both main and eavesdropper channels is available at $S$.

\subsection{FTR Channel Model}
The FTR channel model consists of two fluctuating specular components with random phases plus a diffuse component, and incorporates ground reflections in mmWave channels \cite{Jerez2017the}. The probability distribution function (PDF) and cumulative distribution function (CDF) of the instantaneous signal-to-noise ratio (SNR) over FTR channel are expressed as \cite{Zhang2017New}
\begin{equation}\label{PDF}
{f_i}\left( {{\gamma _i}} \right) = \frac{{m_i^{{m_i}}}}{{\Gamma \left( {{m_i}} \right)}}\sum\limits_{{j_i} = 0}^\infty  {\frac{{K_i^{{j_i}}{d_{{j_i}}}}}{{{j_i}!{j_i}!}}\frac{{\gamma _i^{{j_i}}}}{{{{\left( {2\sigma _i^2} \right)}^{{j_i} + 1}}}}\exp \left( { - \frac{{{\gamma _i}}}{{2\sigma _i^2}}} \right)}  ,
\end{equation}
\begin{equation}\label{CDF}
{F_i}\left( {{\gamma _i}} \right) = \frac{{m_i^{{m_i}}}}{{\Gamma \left( {{m_i}} \right)}}\sum\limits_{{j_i} = 0}^\infty  {\frac{{K_i^{{j_i}}{d_{{j_i}}}}}{{{j_i}!{j_i}!}}\gamma \left( {{j_i} + 1,\frac{{{\gamma _i}}}{{2\sigma _i^2}}} \right)},
\end{equation}
where ${i \in \left\{ {D,E} \right\}}$ represents the main channel or the eavesdropper channel, ${d_{{j_i}}}$ is expressed in terms of the fading parameters ${{m_i}}$, ${{K_i}}$ and ${{\Delta_i}}$, defined in \cite[Eq. (9)]{Zhang2017New}, and $\gamma \left( { \cdot , \cdot } \right)$ is the incomplete gamma function \cite[Eq. (8.350.1)]{Gradshteyn1980In}.

{The performance of mmWave links is also affected by large-scale blockages, such as buildings, in urban areas.
Several past research works, e.g. \cite{Bai2015coverage} and references therein, have pointed out that blockages
result in significant differences between the path loss characteristics of the line-of-sight (LOS) and the non-line-of-sight (NLOS)
components.
In \cite{Bai2015coverage}, the so-called LOS ball blockage model has been considered which approximates general
LOS probability functions as a step function to render mathematical analysis tractable. According to this model, the LOS probability of the link equals to one
within a certain sphere of fixed radius $R_B$ and zero elsewhere. Assuming that the propagation distance, ${r_i}$, lies within this sphere, the average SNR at $D$ or $E$ is given as
\begin{equation}\label{AverSNR}
{{\bar \gamma }_i} = \left( {{{{E_b}} \mathord{\left/
 {\vphantom {{{E_b}} {{N_0}}}} \right.
 \kern-\nulldelimiterspace} {{N_0}}}} \right)2\sigma _i^2\left( {1 + {K_i}} \right)r_i^{ - {\eta _i}},\;\;\;   {i \in \left\{ {D,E} \right\}},
\end{equation}}
where ${{{E_b}}/ {{N_0}}}$ is the energy per bit to the noise power spectral density ratio, {${\eta _i}$ is path-loss exponent,} and $2{{\sigma}_i ^2}$ is the average power of the diffuse component over the FTR fading.

\subsection{Truncation Error}
By truncating \eqref{PDF} up to the first ${{N_i}}+ 1$ terms, the truncation error is given as
\begin{equation}\label{TrunPDF}
{{\hat f}_i}\left( {{\gamma _i}} \right) = \frac{{m_i^{{m_i}}}}{{\Gamma \left( {{m_i}} \right)}}\sum\limits_{{j_i} = 0}^{{N_i}} {\frac{{K_i^{{j_i}}{d_{{j_i}}}\gamma _i^{{j_i}}\exp \left( { - \frac{{{\gamma _i}}}{{2\sigma _i^2}}} \right)}}{{{j_i}!\Gamma \left( {{j_i} + 1} \right){{\left( {2\sigma _i^2} \right)}^{{j_i} + 1}}}}}.
\end{equation}
The truncation error of the area under the ${{f_i}\left( {{\gamma _i}} \right)}$ to the first ${{N_i}}+ 1$ terms is given by
\begin{equation}\label{TrunFunc}
{\varepsilon _i}\left( {{N_i}} \right) \triangleq \int_0^\infty  {{f_i}\left( {{\gamma _i}} \right)d{\gamma _i}}  - \int_0^\infty  {{{\hat f}_i}\left( {{\gamma _i}} \right)d{\gamma _i}}.
\end{equation}
Substituting (\ref{PDF}) and (\ref{TrunPDF}) into (\ref{TrunFunc})
and with the help of \cite[Eq. (8.312.2)]{Gradshteyn1980In}, (\ref{TrunFunc}) can be expressed in closed-form as
\begin{equation}\label{TrunFunc2}
{\varepsilon _i}\left( {{N_i}} \right) = 1 - \frac{{m_i^{{m_i}}}}{{\Gamma \left( {{m_i}} \right)}}\sum\limits_{{j_i} = 0}^{{N_i}} {\frac{{K_i^{{j_i}}{d_{{j_i}}}}}{{{j_i}!}}}.
\end{equation}

\begin{table}[!t]
\renewcommand{\thetable}{\Roman{table}}
\caption{Required terms ${N_i}$ for the truncation error (${{\varepsilon _i} \leq 10^{ - 5}}$) with different channel parameters ${m_i}$, ${K_i}$, and ${{\Delta_i}}$.}
\label{tab1}
\newcommand{\tabincell}[2]{\begin{tabular}{@{}#1@{}}#2\end{tabular}}
\centering          
\begin{tabular}{|c|c|c|c|c|}        
\hline
\hline
FTR Fading Parameters  & ${N_i}$ &  ${\varepsilon _i}$ \\
\hline
\hline
 ${m_i}$=15.5, ${K_i}$=5, ${{\Delta_i}}$=0.4 &  24 & ${6.27 \times {10^{-6}}}$  \\
\hline
 ${m_i}$=8.5, ${K_i}$=5,  ${{\Delta_i}}$=0.35 &  27  & ${6.025 \times {10^{-6}}}$ \\
\hline
 ${m_i}$=25.5, ${K_i}$=3, ${{\Delta_i}}$=0.48 &  16  & ${8.447 \times {10^{-6}}}$ \\
\hline
\hline
\end{tabular}
\end{table}

Table \ref{tab1} depicts the statistic truncation parameter ${{N_i}}$ for different combinations of channel parameters. Note that the maximum required term for accurate ${{N_i}}$ is only 27 in all considered cases. In the realistic propagation environment, the main channel and the eavesdropper channel may have different fading parameters, which results different values of truncation parameters. In this case, we define the truncation parameter $N$ as $N \triangleq \max \{ {N_D},{N_E}\}$.

\section{PHY-security performance analysis over FTR fading channels}\label{se:performance analysis}

\subsection{ASC Analysis}
Recall that the full CSI of both the main and eavesdropper channels is available at $S$, which is called as active eavesdropping \cite{Wang2014Secure}. In such a scenario, $S$ can adapt the achievable secrecy rate to ${R_s}$ such that ${R_s}\leq{C_s}$. Thus, according to \cite{Csiszar1978Broadcast}, the instantaneous secrecy capacity is defined as
\begin{equation}\label{ASC}
{C_s}\left( {{\gamma _D},{\gamma _E}} \right) = \max \left\{ {\ln \left( {1 + {\gamma _D}} \right) - \ln \left( {1 + {\gamma _E}} \right),0} \right\},
\end{equation}
where ${\ln \left( {1 + {\gamma _D}} \right)}$ and ${\ln \left( {1 + {\gamma _E}} \right)}$ are the capacity of the main and eavesdropper channels, respectively. Since both main and eavesdropper channels experience independent fading, the ASC can be given by
\begin{align}\label{ASC1}
  {{\bar C}_s}\left( {{\gamma _D},{\gamma _E}} \right) &=  \int_0^\infty  {\int_0^\infty  {{C_s}\left( {{\gamma _D},{\gamma _E}} \right)f\left( {{\gamma _D},{\gamma _E}} \right)d{\gamma _D}d{\gamma _E}} }  \notag \\
   &= \underbrace {\int_0^\infty  {\ln \left( {1 + {\gamma _D}} \right){f_D}\left( {{\gamma _D}} \right){F_E}\left( {{\gamma _D}} \right)} d{\gamma _D}}_{{I_1}} \notag \\
   &+ \underbrace {\int_0^\infty  {\ln \left( {1 + {\gamma _E}} \right){f_E}\left( {{\gamma _E}} \right){F_D}\left( {{\gamma _E}} \right)} d{\gamma _E}}_{{I_2}} \notag \\
   &- \underbrace {\int_0^\infty  {\ln \left( {1 + {\gamma _E}} \right){f_E}\left( {{\gamma _E}} \right)} d{\gamma _E}}_{{I_3}} , \
\end{align}
where $f\left( {{\gamma _D},{\gamma _E}} \right) = {f_D}\left( {{\gamma _D}} \right){f_E}\left( {{\gamma _E}} \right)$ is the joint pdf of ${{\gamma _D}}$ and ${{\gamma _E}}$.
With the help of (\ref{PDF}), (\ref{CDF}) and (\ref{ASC1}), we can obtain the ASC over FTR fading channels in the following Lemma.

\begin{lemm}
The ASC over FTR fading channels can be expressed as (\ref{Cap}) at the end of next page,
\newcounter{mytempeqncnt}
\begin{figure*}[!b]
\normalsize
\setcounter{mytempeqncnt}{\value{equation}}
\hrulefill
\setcounter{equation}{9}
\begin{align}
  &ASC = \frac{{m_D^{{m_D}}m_E^{{m_E}}}}{{\Gamma \left( {{m_D}} \right)\Gamma \left( {{m_E}} \right)}}\sum\limits_{{j_D} = 0}^\infty  {\sum\limits_{{j_E} = 0}^\infty  {\frac{{K_D^{{j_D}}{d_{{j_D}}}K_E^{{j_E}}{d_{{j_E}}}}}{{{j_D}!{j_E}!}}} } \left( {\frac{{S\left( {{j_D} + 1,{{\left( {2\sigma _D^2} \right)}^{ - 1}}} \right)}}{{{j_D}!{{\left( {2\sigma _D^2} \right)}^{{j_D} + 1}}}} + \frac{{S\left( {{j_E} + 1,{{\left( {2\sigma _E^2} \right)}^{ - 1}}} \right)}}{{{j_E}!{{\left( {2\sigma _E^2} \right)}^{{j_E} + 1}}}}} \right. \hfill \notag \\
 & \left. { - \sum\limits_{n = 0}^{{j_E}} {\frac{{S\left( {{j_D} + n + 1,\frac{{\sigma _D^2 + \sigma _E^2}}{{2\sigma _D^2\sigma _E^2}}} \right)}}{{n!{j_D}!{{\left( {2\sigma _E^2} \right)}^n}{{\left( {2\sigma _D^2} \right)}^{{j_D} + 1}}}}}  - \sum\limits_{n = 0}^{{j_D}} {\frac{{S\left( {{j_E} + n + 1,\frac{{\sigma _D^2 + \sigma _E^2}}{{2\sigma _D^2\sigma _E^2}}} \right)}}{{n!{j_E}!{{\left( {2\sigma _D^2} \right)}^n}{{\left( {2\sigma _E^2} \right)}^{{j_E} + 1}}}}} } \right) - \frac{{m_E^{{m_E}}}}{{\Gamma \left( {{m_E}} \right)}}\sum\limits_{{j_E} = 0}^\infty  {\frac{{K_E^{{j_E}}{d_{{j_E}}}S\left( {{j_E} + 1,{{\left( {2\sigma _E^2} \right)}^{ - 1}}} \right)}}{{{j_E}!{j_E}!{{\left( {2\sigma _E^2} \right)}^{{j_E} + 1}}}}} , \hfill\label{Cap}
\end{align}
\setcounter{equation}{\value{mytempeqncnt}}
\end{figure*}
\addtocounter{equation}{1}
where
\begin{equation}\label{InterS}
S\left( {w,\mu } \right) \triangleq \left( {w - 1} \right)!{e^u}\sum\limits_{k = 1}^w {\frac{{\Gamma \left( { - w + k,\mu } \right)}}{{{\mu ^k}}}}.
\end{equation}
\end{lemm}
\begin{IEEEproof}
Please see Appendix A.
\end{IEEEproof}

Note that \eqref{Cap} is given in terms of only simple functions, which can be efficiently evaluated in common softwares.
\vspace*{-10pt}

\subsection{SOP Analysis}
When $S$ has no information about the eavesdroppers channel, $S$ has no choice but to encode the confidential data into codewords of a constant rate ${R_s}$. If ${R_s} \leqslant {C_s}$, perfect secrecy can be achieved and information theoretic security is compromised. The SOP is defined as the probability that the instantaneous secrecy capacity falls below a target rate, which is an important PHY-security performance metric and widely used to characterize wireless communications. The SOP can be expressed as \cite{Bloch2008Wireless}
\begin{align}\label{SOP}
  SOP &= P\left\{ {{C_s}\left( {{\gamma _D},{\gamma _E}} \right) < {R_s}} \right\} \notag \\
   &= P\left\{ {{\gamma _D} < \Theta {\gamma _E} + \Theta  - 1} \right\} \notag \\
   &= \int_0^\infty  {{F_D}\left( {\Theta {\gamma _E} + \Theta  - 1} \right){f_E}\left( {{\gamma _E}} \right)d{\gamma _E}},
\end{align}
where ${{R_s} \geqslant 0}$ is the target secrecy capacity threshold, and $\Theta  \triangleq {e^{{R_s}}} $. Substituting (\ref{PDF}) and (\ref{CDF}) into (\ref{SOP}), we can obtain the SOP over FTR fading channels in the following Lemma.
\begin{lemm}\label{le:SOP}
The SOP over FTR fading channels can be expressed as
\begin{align}\label{SOP1}
  {\text{SOP}} &= \frac{{m_D^{{m_D}}m_E^{{m_E}}}}{{\Gamma \left( {{m_D}} \right)\Gamma \left( {{m_E}} \right)}}\sum\limits_{{j_D} = 0}^\infty  {\sum\limits_{{j_E} = 0}^\infty  {\frac{{K_D^{{j_D}}{d_{{j_D}}}K_E^{{j_E}}{d_{{j_E}}}}}{{{j_D}!{j_E}!}}} }  \notag \\
   &\times \left( {1 - \sum\limits_{n = 0}^{{j_D}} {\sum\limits_{q = 0}^n {\left( {\begin{array}{*{20}{c}}
  n \\
  q
\end{array}} \right)\frac{1}{{n!{j_E}!}}\exp \left( { - \frac{{\Theta  - 1}}{{2\sigma _D^2}}} \right)} } } \right. \notag \\
   &\times \left. {\frac{{{\Theta ^q}{{\left( {\Theta  - 1} \right)}^{n - q}}\Gamma \left( {{j_E} + 1 + q} \right){{\left( {2\sigma _E^2} \right)}^q}}}{{{{\left( {1 + \frac{{\sigma _E^2\Theta }}{{\sigma _D^2}}} \right)}^{{j_E} + q + 1}}{{\left( {2\sigma _D^2} \right)}^n}}}} \right).
\end{align}

\end{lemm}
\begin{IEEEproof}
Please see Appendix B.
\end{IEEEproof}

By adopting a similar method in \cite{Lei2015On}, we derive the lower bound of the SOP based on (\ref{SOP}) as
\begin{align}\label{SOPL}
{SOP}^{L} = P\left\{ {{\gamma _D} < \Theta {\gamma _E}} \right\} \le {SOP}.
\end{align}

Substituting (\ref{PDF}) and (\ref{CDF}) into (\ref{SOPL}), the lower bound of the SOP over FTR fading channels is derived in the following Lemma.

\begin{lemm}\label{le:SOPL}
The lower bound of the SOP over FTR fading channels can be expressed as
\begin{align}\label{SOPL1}
  &SO{P^L} = \frac{{m_D^{{m_D}}m_E^{{m_E}}}}{{\Gamma \left( {{m_D}} \right)\Gamma \left( {{m_E}} \right)}}\sum\limits_{{j_D} = 0}^\infty  {\sum\limits_{{j_E} = 0}^\infty  {\frac{{K_D^{{j_D}}{d_{{j_D}}}K_E^{{j_E}}{d_{{j_E}}}}}{{{j_D}!{j_E}!}}} }  \notag \\
   &\times \frac{{{{\left( {\rho \eta } \right)}^{{j_E}}}{\Theta ^{{j_D} + 1}}}}{{{{\left( {\Theta  + \rho \eta } \right)}^{{j_D} + {j_E} + 1}}}}\sum\limits_{k = 0}^{{j_E}} {{{\left( {\frac{\Theta }{{\rho \eta }}} \right)}^k}\frac{{\left( {{j_D} + {j_E} + 1} \right)!}}{{\left( {{j_D} + 1 + k} \right)!\left( {{j_E} - k} \right)!}}},  \hfill
\end{align}
where $\rho  \triangleq \frac{{{{\bar \gamma }_D}}}{{{{\bar \gamma }_E}}} = \frac{{\sigma _D^2\left( {{K_D} + 1} \right)}}{{\sigma _E^2\left( {{K_E} + 1} \right)}}$ and $\eta  \triangleq \frac{{{K_E} + 1}}{{{K_D} + 1}}$.
\end{lemm}
\begin{IEEEproof}
Please see Appendix C.
\end{IEEEproof}

\subsection{SPSC Analysis}
The probability of SPSC, which is a fundamental benchmark in secure communications, can be obtained by \cite{Bloch2008Wireless}
\begin{align}\label{SPSC0}
  SPSC &= P\left\{ {{C_s}\left( {{\gamma _D},{\gamma _E}} \right) > 0} \right\}  = P\left\{ {{\gamma _D} > {\gamma _E}} \right\} \notag \\
   &= 1 - SOP_{{R_s}=0}^L.
\end{align}
Therefore, we can obtain SPSC by substituting (\ref{SOPL1}) into (\ref{SPSC0}) and setting $\Theta  = {e^{{R_s}}} = 1$.

\section{Numerical Results}\label{se:numerical_result}
In this section, we present some plots that illustrate the ASC, SOP and SPSC of mmWave communications over FTR fading channels with. For the Monte Carlo simulation, $10^6$ realizations of FTR fading channels are generated to validate the analytical expressions derived in previous sections {and the propagation distance $r_i$ is normalized to 1 km.}

\begin{figure}[t]
\centering
\includegraphics[scale=0.5]{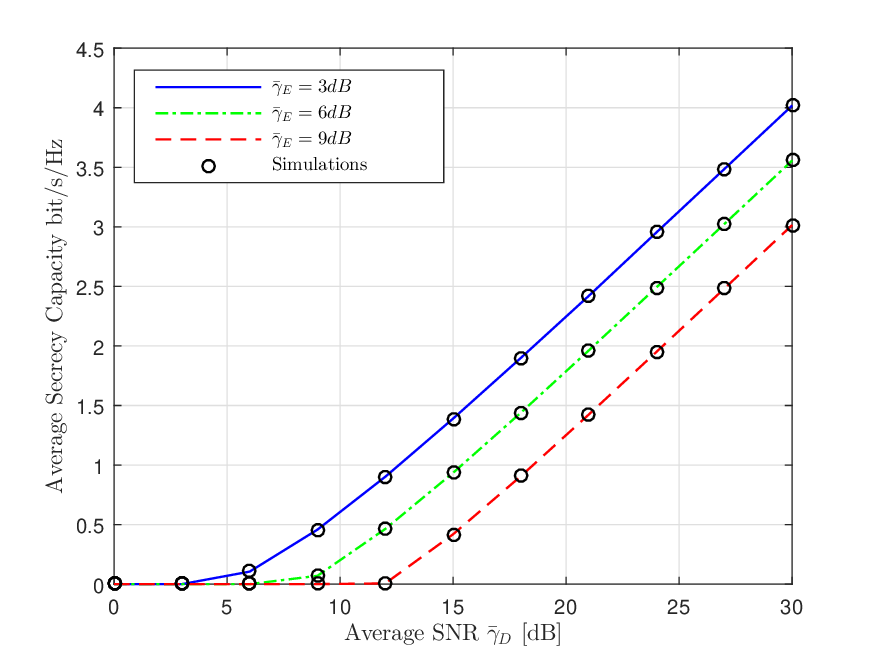}
\caption{ASC over FTR fading channels against ${{\bar \gamma }_D}$ for different values of ${{\bar \gamma }_E}$ ($K_D=15, K_E=5$, $m_D=5.5, m_E=8.5$, ${\Delta}_D=0.4$, and $ {\Delta}_E=0.35$). }
\label{ASCE}
\end{figure}

The ASC as a function of ${{\bar \gamma }_D}$ in dB is depicted in Fig. \ref{ASCE} for ${{\bar \gamma }_E} = 3, 6, 9$ dB. The outputs of a Monte Carlo simulator are shown to exactly match with the analytical results, which validates our derived results. As expected, the performance of ASC improves with increasing ${{\bar \gamma }_D}$ or decreasing ${{\bar \gamma }_E}$. Note that the ASC will fall to zero if the average SNR of the main channel is smaller than the eavesdropper channel (${{\bar \gamma }_D} < {{\bar \gamma }_E}$ ), which is consistent with (\ref{ASC}).

\begin{figure}[t]
\centering
\includegraphics[scale=0.5]{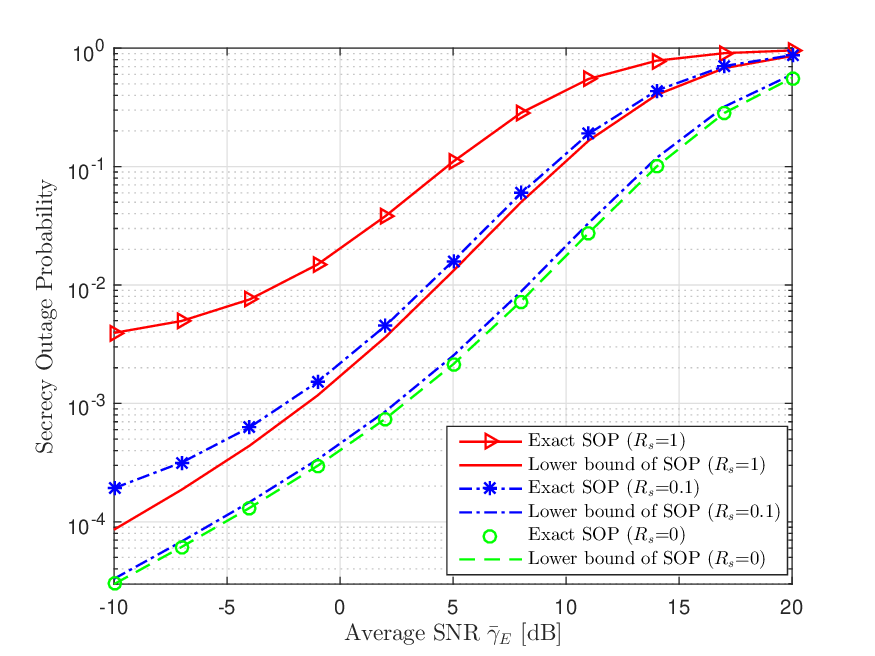}
\caption{SOP over FTR fading channels against ${{\bar \gamma }_E}$ for different values of ${R_s}$ ($K_D=15, K_E=5$, $m_D=5.5, m_E=8.5$, ${\Delta}_D=0.4$, ${\Delta}_E=0.35$, and ${{\bar \gamma }_D}=15$dB ). }
\label{SOP_SOPL}
\end{figure}

In Fig. \ref{SOP_SOPL}, we portray the exact and the lower bound of SOP as a function of the average SNR of the eavesdropper channel ${{\bar \gamma }_E}$. The high-SNRs of ${{\bar \gamma }_E}$ make the lower bound of the SOP sufficiently tight with the exact SOP. It is clear that the lower bound of SOP becomes accurate as the value of ${R _s}$ decreases. Moreover, it can be observed that the SOP performance of the considered system is improved by decreasing the values of ${{\bar \gamma }_E}$, which is consistent with the results presented in Lemma \ref{le:SOP} and Lemma \ref{le:SOPL}.

\begin{figure}[t]
\centering
\includegraphics[scale=0.5]{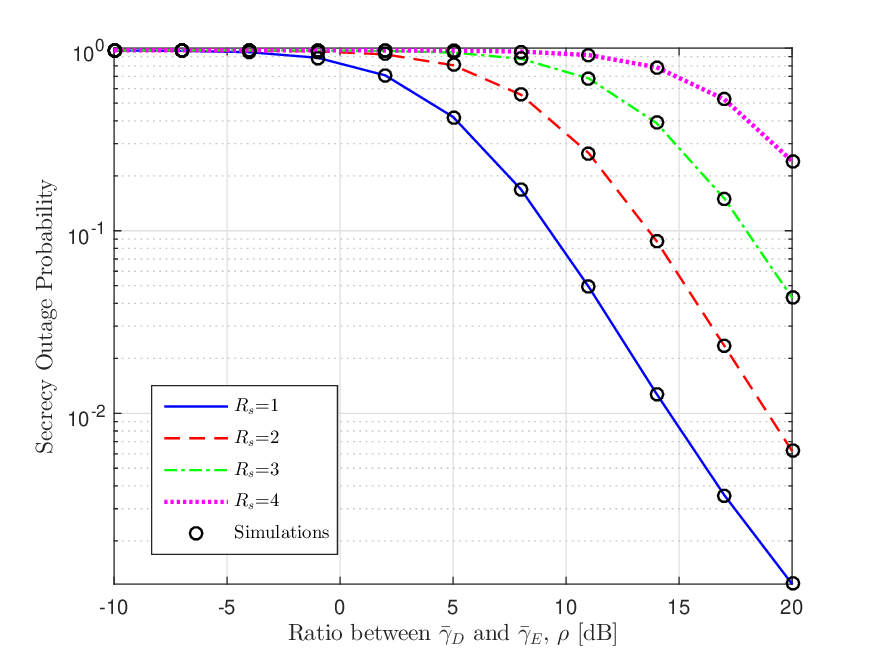}
\caption{SOP over FTR fading channels against ${\rho}$ for different values of ${R _{s}}$ ($K_D=K_E=8$, $m_D=m_E=5.5$, and ${\Delta}_D={\Delta}_E=0.4$). }
\label{SOP_Q}
\end{figure}

Fig. \ref{SOP_Q} investigates the impact of the ratios between ${{\bar \gamma }_D}$ and $ {{\bar \gamma }_E}$, $\rho$, on the SOP performance. The achievable secrecy rates ${R_s}$ are considered (${R_s} = 1, 2, 3, 4 $ bit/s/Hz). Intuitively, as $\rho$ become large, the main channel is much better than the eavesdropper channel and the SOP becomes decreasingly substantial. In addition, smaller ${R_s}$ can obtain smaller SOP, which is consistent with the results presented in Lemma \ref{le:SOP}.

\begin{figure}[htbp]
\centering
\includegraphics[scale=0.5]{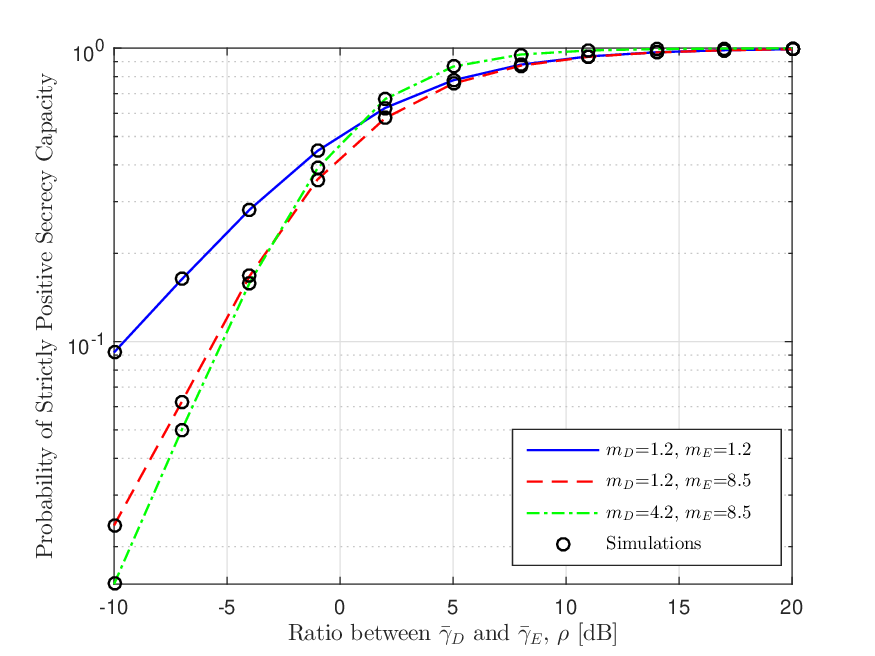}
\caption{SPSC over FTR fading channels against ${\rho}$ for different values of $m_D$ and $m_E$ ($K_D=K_E=8$, ${\Delta}_D={\Delta}_E=0.3$, and ${R _{s}} = 0$). }
\label{0SPSC}
\end{figure}

Fig. \ref{0SPSC} illustrates the effect of shadowing on the SPSC performance of mmWave communications over FTR fading channels. As can be readily observed, the light shadowing (small values of $m$) in eavesdropper channel will increase the SPSC. Furthermore, in the moderate- and high-$\rho$ regime, increasing the shadowing effect of the main channel $m_D$
can increase the SPSC performance, which is not observed in the very low-$\rho$ regime.

\vspace*{-2pt}
\section{Conclusion}\label{se:conclusion}
In this correspondence, we investigate the PHY-security performance of mmWave communications over FTR fading channels. We derive analytical expressions for the ASC, SOP and SPSC in terms of simple functions, which can quickly and steadily converge with only a few of $N$ terms to obtain a desired accuracy. Note that derived results can reduce to many pioneering works, since the FTR includes Rayleigh, Rician, and Nakagami-$m$ as special cases. Our analysis validates that the performance of the considered system can be improved with increasing the average SNR of the main channel or decreasing the average SNR of the eavesdropper channel. Moreover, the light shadowing (small values of $m$) in eavesdropper channel will increase the SPSC. As for current and future directions, it is of interest to investigate the PHY-security performance of mmWave communications by considering more practical channel and system features, such as blockages, interference, and multi-antenna.

\section{APPENDIX}\label{se:APPENDIX}

\subsection{Proof of Lemma 1}
For the natural number ${j_i}$, the gamma function $\Gamma \left(  \cdot  \right)$ can be expressed as $\Gamma \left( {{j_i} + 1} \right) = {j_i}!$ \cite[Eq. (8.339.1)]{Gradshteyn1980In}. Then, substituting (\ref{PDF}) and (\ref{CDF}) into (\ref{ASC1}), ${I_1}$ can be expressed as
\begin{align}\label{I1}
  &{I_1} = \frac{{m_D^{{m_D}}}}{{\Gamma \left( {{m_D}} \right)}}\frac{{m_E^{{m_E}}}}{{\Gamma \left( {{m_E}} \right)}}\sum\limits_{{j_D} = 0}^\infty  {\sum\limits_{{j_E} = 0}^\infty  {\frac{{K_D^{{j_D}}{d_{{j_D}}}K_E^{{j_E}}{d_{{j_E}}}}}{{{j_D}!{j_E}!{j_D}!{j_E}!{{\left( {2\sigma _D^2} \right)}^{{j_D} + 1}}}}} }  \hfill\notag \\
   &\times \underbrace {\int_0^\infty  {\ln \left( {1 + {\gamma _D}} \right)\gamma _D^{{j_D}}{e^{ - \frac{{{\gamma _D}}}{{2\sigma _D^2}}}}\gamma \left( {{j_E} + 1,\frac{{{\gamma _D}}}{{2\sigma _E^2}}} \right)} d{\gamma _D}}_{{A_1}},
\end{align}
In order to solve the inner integral ${A_1}$, with the help of \cite[Eq. (8.354.1)]{Gradshteyn1980In}, we have
\begin{align}\label{InGamma}
\gamma \left( {{j_E} + 1,\frac{{{\gamma _D}}}{{2\sigma _E^2}}} \right) = {j_E}!\left( {1 - {e^{ - \frac{{{\gamma _D}}}{{2\sigma _E^2}}}}\sum\limits_{n = 0}^{{j_E}} {\frac{1}{{n!}}{{\left( {\frac{{{\gamma _D}}}{{2\sigma _E^2}}} \right)}^n}} } \right).
\end{align}
Substituting (\ref{InGamma}) into ${A_1}$ and formulating the integral as $S\left( {w,\mu } \right) \triangleq \int_0^\infty  {\ln \left( {1 + t} \right){t^{w - 1}}{e^{ - \mu t}}} dt$, we can obtain
\begin{align}\label{A1}
  &{A_1} = {j_E}!S\left( {{j_D} + 1,\frac{1}{{2\sigma _D^2}}} \right) \notag \\
   &- {j_E}!\sum\limits_{n = 0}^{{j_E}} {\frac{1}{{n!}}{{\left( {\frac{1}{{2\sigma _E^2}}} \right)}^n}} S\left( {{j_D} + n + 1,\frac{{\sigma _D^2 + \sigma _E^2}}{{2\sigma _D^2\sigma _E^2}}} \right).
\end{align}
Since $w$ is a natural number in the integral $S\left( {w,\mu } \right)$, we can have (\ref{InterS}) as in \cite{Alouini1999Capacity}.
Substituting (\ref{A1}) and (\ref{InterS}) into (\ref{I1}), and after a simple transformation of the variables, ${I_1}$ is given as
\begin{align}\label{I1-1}
  &{I_1} = \frac{{m_D^{{m_D}}m_E^{{m_E}}}}{{\Gamma \left( {{m_D}} \right)\Gamma \left( {{m_E}} \right)}}\sum\limits_{{j_D} = 0}^\infty  {\sum\limits_{{j_E} = 0}^\infty  {\frac{{K_D^{{j_D}}{d_{{j_D}}}K_E^{{j_E}}{d_{{j_E}}}}}{{{j_D}!{j_E}!{j_D}!{{\left( {2\sigma _D^2} \right)}^{{j_D} + 1}}}}} }  \hfill \\
  & \times \left( {S\left( {{j_D} + 1,\frac{1}{{2\sigma _D^2}}} \right) - \sum\limits_{n = 0}^{{j_E}} {\frac{{S\left( {{j_D} + n + 1,\frac{{\sigma _D^2 + \sigma _E^2}}{{2\sigma _D^2\sigma _E^2}}} \right)}}{{n!{{\left( {2\sigma _E^2} \right)}^n}}}} } \right). \notag
\end{align}
Following similar steps, we can obtain ${I_2}$ and ${I_3}$ as
\begin{align}\label{I2}
  &{I_2} = \frac{{m_D^{{m_D}}m_E^{{m_E}}}}{{\Gamma \left( {{m_D}} \right)\Gamma \left( {{m_E}} \right)}}\sum\limits_{{j_D} = 0}^\infty  {\sum\limits_{{j_E} = 0}^\infty  {\frac{{K_D^{{j_D}}{d_{{j_D}}}K_E^{{j_E}}{d_{{j_E}}}}}{{{j_D}!{j_E}!{j_E}!{{\left( {2\sigma _E^2} \right)}^{{j_E} + 1}}}}} }  \hfill\\
   &\times \left( {S\left( {{j_E} + 1,\frac{1}{{2\sigma _E^2}}} \right) - \sum\limits_{n = 0}^{{j_D}} {\frac{{S\left( {{j_E} + n + 1,\frac{{\sigma _D^2 + \sigma _E^2}}{{2\sigma _D^2\sigma _E^2}}} \right)}}{{n!{{\left( {2\sigma _D^2} \right)}^n}}}} } \right). \hfill \notag
\end{align}
\begin{equation}\label{I3}
{I_3} = \frac{{m_E^{{m_E}}}}{{\Gamma \left( {{m_E}} \right)}}\sum\limits_{{j_E} = 0}^\infty  {\frac{{K_E^{{j_E}}{d_{{j_E}}}} S\left( {{j_E} + 1,{({2\sigma _E^2})^{-1}}}\right)}{{{j_E}!{j_E}!{{\left( {2\sigma _E^2} \right)}^{{j_E} + 1}}}}} .
\end{equation}
Then, we can obtain (\ref{Cap}) by combining (\ref{I1-1}), (\ref{I2}) and (\ref{I3}).

\vspace{-0.1cm}
\subsection{Proof of Lemma 2}
Substituting (\ref{PDF}) and (\ref{CDF}) into (\ref{SOP}), we can obtain
\begin{align}\label{S1}
  &{\text{SOP}} = \frac{{m_D^{{m_D}}m_E^{{m_E}}}}{{\Gamma \left( {{m_D}} \right)\Gamma \left( {{m_E}} \right)}}\sum\limits_{{j_D} = 0}^\infty  {\sum\limits_{{j_E} = 0}^\infty  {\frac{{K_D^{{j_D}}{d_{{j_D}}}K_E^{{j_E}}{d_{{j_E}}}}}{{{j_D}!{j_E}!{j_D}!{j_E}!{{\left( {2\sigma _E^2} \right)}^{{j_E} + 1}}}}} }  \hfill \notag\\
  & \times \underbrace {\int_0^\infty  {\gamma _E^{{j_E}}{e^{ - \frac{{{\gamma _E}}}{{2\sigma _E^2}}}}\gamma \left( {{j_D} + 1,\frac{{\Theta {\gamma _E} + \Theta  - 1}}{{2\sigma _D^2}}} \right)d{\gamma _E}} }_{{I_4}}. \hfill
\end{align}
With the help of \cite[Eq. (8.354.1)]{Gradshteyn1980In}, ${I_4}$ can be expressed as
\begin{align}\label{S1-1}
 & {I_4} = {j_D}!\underbrace {\int_0^\infty  {\gamma _E^{{j_E}}{e^{ - \frac{{{\gamma _E}}}{{2\sigma _E^2}}}}d{\gamma _E}} }_{{I_5}} - {j_D}!\sum\limits_{n = 0}^{{j_D}} {\frac{1}{{n!}}{{\left( {\frac{1}{{2\sigma _D^2}}} \right)}^n}{e^{\frac{{1 - \Theta }}{{2\sigma _D^2}}}}} \notag  \\
 &  \times \underbrace {\int_0^\infty  {\gamma _E^{{j_E}}{e^{ - \frac{{{\gamma _E}}}{{2\sigma _E^2}} - \frac{{\Theta {\gamma _E}}}{{2\sigma _D^2}}}}{{\left( {\Theta {\gamma _E} + \Theta  - 1} \right)}^n}d{\gamma _E}} }_{{I_6}}.
\end{align}
Using \cite[Eq. (3.326)]{Gradshteyn1980In} and \cite[Eq. (1.111)]{Gradshteyn1980In}, we have
\begin{equation}\label{S2}
{I_5} = \Gamma \left( {{j_E} + 1} \right){\left( {2\sigma _E^2} \right)^{{j_E} + 1}},
\end{equation}
\begin{align}\label{S3}
  {I_6} = \sum\limits_{q = 0}^n {\left( {\begin{array}{*{20}{c}}
  n \\
  q
\end{array}} \right)} \frac{{{\Theta ^q}{{\left( {\Theta  - 1} \right)}^{n - q}}\Gamma \left( {{j_E} + 1 + q} \right)}}{{{{\left( {\frac{1}{{2\sigma _E^2}} + \frac{\Theta }{{2\sigma _D^2}}} \right)}^{{j_E} + q + 1}}}}. \hfill
\end{align}
The proof concludes by combining (\ref{S1}), (\ref{S1-1}), (\ref{S2}), and (\ref{S3}).

\vspace{-0.3cm}
\subsection{Proof of Lemma 3}
Substituting (\ref{PDF}) and (\ref{CDF}) into (\ref{SOPL}), we can obtain
\begin{align}\label{S4}
  &SO{P^L} = \frac{{m_D^{{m_D}}m_E^{{m_E}}}}{{\Gamma \left( {{m_D}} \right)\Gamma \left( {{m_E}} \right)}}\sum\limits_{{j_D} = 0}^\infty  {\sum\limits_{{j_E} = 0}^\infty  {\frac{{K_D^{{j_D}}{d_{{j_D}}}K_E^{{j_E}}{d_{{j_E}}}}}{{{j_D}!{j_E}!{j_D}!{j_E}!{{\left( {2\sigma _E^2} \right)}^{{j_E} + 1}}}}} }  \notag\\
   &\times \underbrace {\int_0^\infty  {\gamma _E^{{j_E}}\exp \left( { - \frac{{{\gamma _E}}}{{2\sigma _E^2}}} \right)\gamma \left( {{j_D} + 1,\frac{{\Theta {\gamma _E}}}{{2\sigma _D^2}}} \right)d{\gamma _E}} }_{{I_7}}. \hfill
\end{align}
With the help of \cite[Eq. (6.455.2)]{Gradshteyn1980In}, ${I_7}$ can be expressed as
\begin{align}\label{S4-1}
  {I_7} &= \frac{{\Gamma \left( {{j_D} + {j_E} + 2} \right)}}{{\left( {{j_D} + 1} \right)}}{\left( {\frac{\Theta }{{2\sigma _D^2}}} \right)^{{j_D} + 1}}{\left( {\frac{\Theta }{{2\sigma _D^2}} + \frac{1}{{2\sigma _E^2}}} \right)^{ - \left( {{j_D} + {j_E} + 2} \right)}} \notag \\
   &\times {}_2{F_1}\left( {1,{j_D} + {j_E} + 2;{j_D} + 2;\frac{{\sigma _E^2\Theta }}{{\sigma _E^2\Theta  + \sigma _D^2}}} \right), \hfill
\end{align}
where ${}_2{F_1}\left( { \cdot , \cdot ; \cdot ; \cdot } \right)$ is the Gauss hypergeometric function \cite[Eq. (9.14)]{Gradshteyn1980In}. Using \cite[Eq. (7.3.1.129)]{prudnikov1986integrals3}, the proof concludes by combining (\ref{S4-1}) and (\ref{S4}) with some simplifications.

\bibliographystyle{IEEEtran}
\bibliography{IEEEabrv,Ref}
\end{document}